\documentclass[aps,prl,twocolumn,superscriptaddress]{revtex4-1}
\usepackage{graphicx}
\usepackage{amsmath}
\usepackage{amssymb}
\usepackage{color}
\newcommand{\rr}{\color{black}}

\begin{document}
\title{Lowest-energy Moir\'e Band Formed by Dirac Zero Modes in Twisted Bilayer Graphene}

\author{Long Zhang}
\email{longzhang@ucas.ac.cn}
\affiliation{Kavli Institute for Theoretical Sciences and CAS Center for Excellence in Topological Quantum Computation, University of Chinese Academy of Sciences, Beijing 100190, China}
\affiliation{Physical Science Laboratory, Huairou National Comprehensive Science Center, Beijing 101400, China}

\date{\today}

\begin{abstract}
An unconventional insulating phase and a superconducting phase were recently discovered in the twisted bilayer graphene [Y. Cao et al, Nature {\bf 556}, 80; {\bf 556}, 43 (2018)], but the relevant low-energy electronic states have not been clearly identified yet. In this work, I show that the interlayer hopping induces a spatially modulated Dirac mass term in the continuum Hamiltonian, and leads to a low-energy band formed by Dirac zero modes in the moir\'e superlattice. This moir\'e band becomes extremely flat and thus strongly correlated as the Dirac velocity vanishes at the magic angle, and enters a quantum disordered Mott insulating phase at $1/4$ and $3/4$ filling, i.e., $\pm 2$ excess electrons per moir\'e supercell, which may account for the insulating phase discovered in experiments.
\end{abstract}
\maketitle

\begin{figure}[b]
\includegraphics[width=0.5\textwidth]{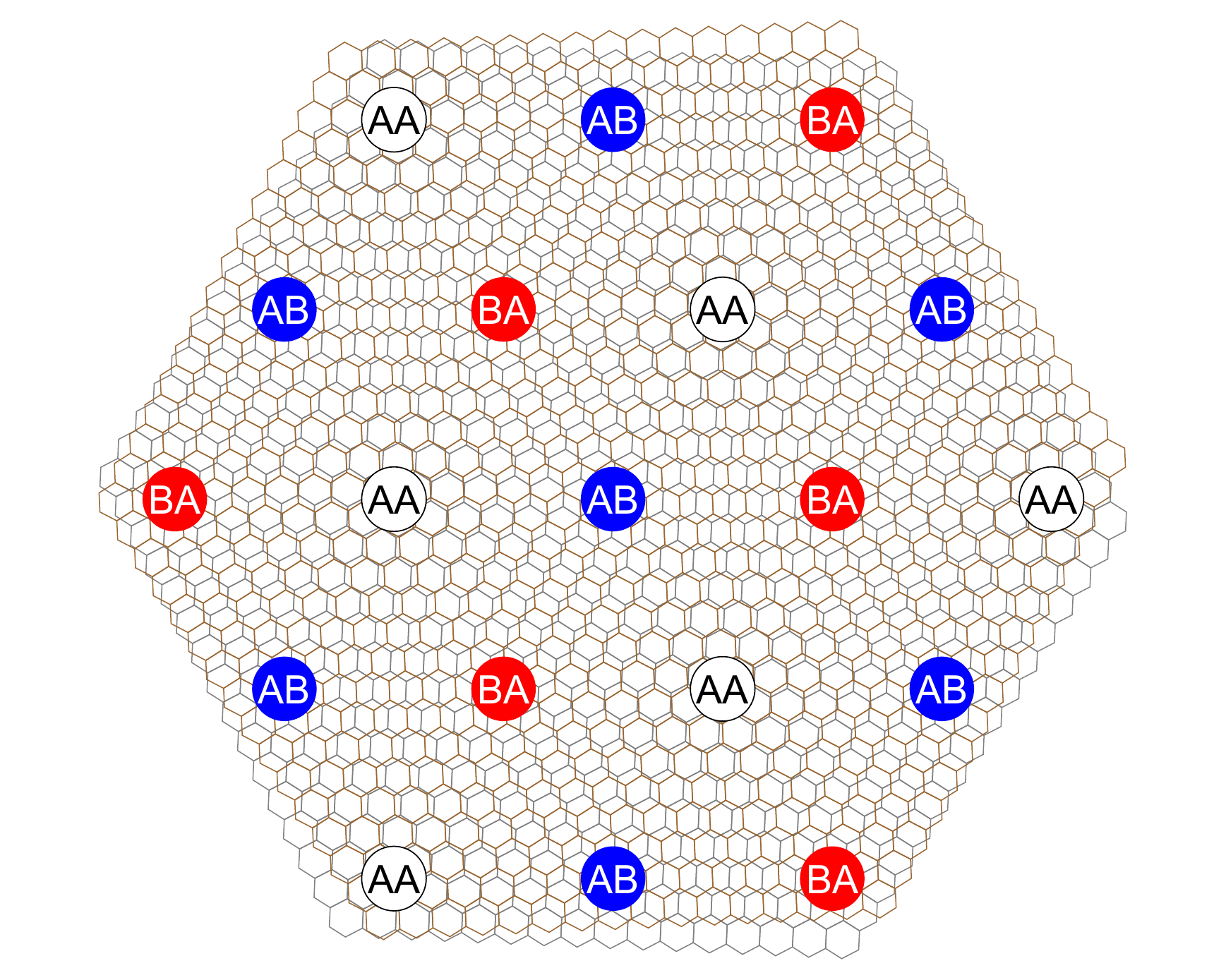}
\caption{Illustration of the moir\'e superstructure in the twisted bilayer graphene. The two layers are rotated relatively by a small angle, and show a spatially modulated stacking pattern with AB, BA and AA stacking regions appearing in each supercell. The Dirac mass term $H'$ in Eq. (\ref{eq:hperptr}) induced by interlayer hopping has a vortex (antivortex) structure around each AB (BA) stacking region, which hosts a Dirac zero mode with chirality $+1$ ($-1$). These regions form an emergent honeycomb lattice.}
\label{fig:lattice}
\end{figure}

\emph{Introduction.---}The recent discovery of an unconventional insulating phase and an adjacent superconducting (SC) phase in the twisted bilayer graphene (TBG) \cite{Cao2018, Cao2018a} has triggered great excitement \cite{Ramires2018, Xu2018, Yuan2018, Po2018, Roy2018, Liu2018a, Koshino2018, Kang2018, Gonzalez2019}. The two layers of graphene are rotated relatively by an angle $\theta$. The bilayer forms a moir\'e pattern (see Fig. \ref{fig:lattice}), i.e., a superstructure with a large superlattice constant $\lambda = a_{0}/2\sin(\theta/2)$, in which $a_{0}$ is the lattice constant of the monolayer graphene. The bilayer has a spatially modulated stacking patter, varying from AB, BA, to AA stacking within each supercell. The rotation angle $\theta \simeq 1.08^{\circ}$ in the experiments \cite{Cao2018, Cao2018a} is the largest one of a discrete set of ``magic angles'', at which the Fermi velocity of the massless Dirac cone in graphene is suppressed to zero and the low-energy band becomes nearly flat \cite{LopesDosSantos2007, TramblydeLaissardiere2010, Bistritzer2011}. An insulating phase sets in at low temperatures in the TBG samples electrically gated away from the charge neutral point with $\pm 2$ excess electrons per supercell, and an SC phase emerges in samples slightly doped away from this insulating phase. The maximum SC transition temperature is $1.7$ K.

The key to understanding these unconventional phases in TBG is to first identify the lowest-energy electronic states around the charge neutral point in the moir\'e superstructure. Some recent theoretical works studied effective models taken either from phenomenological arguments or based on global symmetry analysis \cite{Po2018}; however, microscopic approaches are more desirable (cf. e.g., Refs. \cite{Ramires2018, Efimkin2018, Yuan2018, Roy2018}). This is the main goal of this work.

The microscopic model of the moir\'e band has been studied intensively with tight-binding models \cite{TramblydeLaissardiere2010} and continuum models \cite{LopesDosSantos2007, Bistritzer2011}. Based on the microscopic modeling of the interlayer hopping in the moir\'e superstructure, it has been shown \cite{LopesDosSantos2007, Bistritzer2011} that the Dirac cone band structure of the monolayer graphene survives in the TBG, but the Dirac velocity $v_{F}$ is strongly suppressed at small twist angles due to the interlayer hopping. Moreover, it has been shown \cite{TramblydeLaissardiere2010, Bistritzer2011} that the renormalized Dirac velocity $\tilde{v}_{F}$ \footnote{The renormalized Dirac velocity is denoted by $\tilde{v}_{F}$ instead of $v_{F}^{*}$ in Ref. \cite{Bistritzer2011}. The superscript $^{*}$ is reserved for complex conjugate.} vanishes at the magic angles, and the low-energy band becomes nearly flat. The nearly flat band at magic angles have been confirmed in experiments \cite{Luican2011, Yin2015a, Li2017d, Cao2018}.

In this work, I take the continuum model of the moir\'e band \cite{Bistritzer2011} as the starting point. I first show that a term in the interlayer hopping Hamiltonian, which has been overlooked so far due to the small twist angle, turns out to be a spatially modulated Dirac mass term. This mass term has vortex (antivortex) structures in AB (BA) stacking regions, respectively, which form an emergent honeycomb lattice in the moir\'e superstructure (see Fig. \ref{fig:lattice}). Each vortex (antivortex) traps a Dirac zero mode with chirality $+1$ ($-1$) for topological reason, which becomes extremely localized around the vortex center when the Dirac velocity $\tilde{v}_{F}$ vanishes at the magic angles. There are eight zero modes in each supercell if the spin and the valley degeneracies are taken into account. These Dirac zero modes should be responsible for the unconventional phases at low temperatures discovered in experiments.

The spatial overlap of these zero modes leads to a low-energy band around the charge neutral point. The effective Hamiltonian is an SU(4)-symmetric Hubbard model on the emergent honeycomb lattice, where the four flavors come from the spin and the valley degeneracies. At $1/4$ or $3/4$ filling, which corresponds to $\pm 2$ electrons per supercell away from the charge neutral point in TBG, the SU(4) Hubbard model on the honeycomb lattice forms a quantum disordered Mott insulating phase at the ground state \cite{Corboz2012}, which may account for the insulating phase in experiments. This model should be a good starting point for further theoretical study.

\emph{Revisit to moir\'e band theory.---}Let us start by recapitulating the moir\'e band theory of TBG in the continuum limit \cite{LopesDosSantos2007, Bistritzer2011}. The electron hopping within each layer leads to the well-known massless Dirac cone band structure. In the continuum limit, the intralayer hopping Hamiltonian expanded around one of the Dirac cones at momentum $\vec{K}=(4\pi/3a_{0}, 0)$ is given by
\begin{equation} \label{eq:hintra}
H_{0}=\sum_{\vec{k}}
\Psi_{\vec{k}}^{\dag}
\begin{pmatrix}
h_{\vec{k}}(\theta/2) & \\
& h_{\vec{k}}(-\theta/2)
\end{pmatrix}
\Psi_{\vec{k}},
\end{equation}
in which $\Psi_{\vec{k}}=(\psi_{1\vec{k}}, \psi_{2\vec{k}})^{T}$, and $\psi_{1(2),\vec{k}}$ is the two-component Dirac spinor in the top (bottom) layer. An AB stacking pattern is assumed at the origin for convenience, and the energy spectrum does not depend on this choice \cite{Bistritzer2011}. The momentum $\vec{k}$ is measured from $\vec{K}$. The Dirac Hamiltonian $h_{\vec{k}}(\pm \theta/2)$ is given by
\begin{equation}
\begin{split}
h_{\vec{k}}(\pm\theta/2)=&-v_{F}k
\begin{pmatrix}
0 & e^{i(\theta_{k}\mp \theta/2)} \\
e^{-i(\theta_{k}\mp \theta/2)} & 0
\end{pmatrix}\\
=&-v_{F}e^{\mp i\theta\sigma_{z}/4}\vec{k}\cdot \vec{\sigma}^{*}e^{\pm i\theta\sigma_{z}/4},
\end{split}
\end{equation}
in which $\pm\theta/2$ are the rotation angles of the two layers. $k$ and $\theta_{k}$ are the magnitude and the polar angle of $\vec{k}$, respectively. $\vec{\sigma}=(\sigma_{x},\sigma_{y})$ are the Pauli matrices acting on the Dirac spinors, and $\vec{\sigma}^{*}$ denotes the complex conjugate.

The momentum transfer induced by the interlayer hopping is small due to the large interlayer distance and the smooth interlayer hopping amplitude \cite{Bistritzer2011}, thus the interlayer hopping predominantly takes place between the same valleys of the two layers, thus the valley degeneracy is approximately maintained. Therefore, we only focus on the $\vec{K}$ valley, and similar results can be easily derived for the other valley at $\vec{K}'=(-4\pi/3a_{0},0)$.

The interlayer hopping Hamiltonian projected onto the $\vec{K}$ valley states is spatially modulated due to the moir\'e superstructure \cite{Bistritzer2011},
\begin{equation} \label{eq:hinter}
H_{\perp} = \sum_{\vec{r}}\psi_{1\vec{r}}^{\dag}T(\vec{r})\psi_{2\vec{r}}+\mathrm{h.c.}
\end{equation}
Here the real-space representation is adopted for later convenience. The hopping matrix
\begin{equation}
T(\vec{r})=w\sum_{j=1}^{3}e^{-i\vec{q}_{j}\cdot \vec{r}}T_{j},
\end{equation}
in which the interlayer hopping strength $w\simeq 110$ meV. $\vec{q}_{j}$'s are the interlayer momentum transfers. They have the same magnitude $k_{\theta}\equiv |\vec{q}_{j}|= 8\pi\sin(\theta/2)/3a_{0}$, and are along $(0,-1)$, $(\sqrt{3}/2,1/2)$, and $(-\sqrt{3}/2,1/2)$ directions, respectively. The $T_{j}$ matrices are
\begin{equation}
T_{1}=
\begin{pmatrix}
1 & 1 \\
1 & 1
\end{pmatrix}, \quad
T_{2}= T_{3}^{*}=
\begin{pmatrix}
e^{-i\phi} & 1 \\
e^{i\phi} & e^{-i\phi}
\end{pmatrix},
\end{equation}
in which $\phi = 2\pi /3$.

It has been shown \cite{Bistritzer2011} that if the $\theta$ dependence in $h(\pm\theta/2)$ is neglected, the interlayer hopping does not change the form of the Dirac Hamiltonian in Eq. (\ref{eq:hintra}), but the Dirac velocity $v_{F}$ is strongly renormalized to $\tilde{v}_{F}$ at small twist angles. More remarkably, $\tilde{v}_{F}$ vanishes at the magic angles, which results into a nearly flat band around the charge neutral point.

In this work, I shall show that the $\theta$ dependence in $h(\pm\theta/2)$ has a significant impact on the low-energy electronic states. Let us first absorb the $\theta$ dependence of $h(\pm\theta/2)$ into the electron operators by the following unitary transformation,
\begin{equation}
\Psi_{\vec{r}}\rightarrow e^{i\theta\sigma_{z}\otimes \tau_{z}/4}\Psi_{\vec{r}}= (e^{i\theta\sigma_{z}/4}\psi_{1\vec{r}}, e^{-i\theta\sigma_{z}/4}\psi_{2\vec{r}})^{T},
\end{equation}
in which the Pauli matrix $\tau_{z}$ acts on the bilayer index. Then the interlayer hopping term $H_{\perp}$ is transformed into
\begin{equation} \label{eq:hperptr}
H_{\perp}\rightarrow H_{\perp} + \sum_{\vec{r}}\Psi_{\vec{r}}^{\dag}\sigma_{z}\otimes \tau(\vec{r})\Psi_{\vec{r}} \equiv H_{\perp}+H',
\end{equation}
in which
\begin{equation}
\begin{split}
\tau(\vec{r})=& \frac{1}{2}w\theta[(\sin\vec{q}_{1}\cdot \vec{r}+\sin (\vec{q}_{2}\cdot \vec{r}+\phi)+\sin (\vec{q}_{3}\cdot \vec{r}-\phi))\tau_{x}\\
&-(\cos\vec{q}_{1}\cdot \vec{r}+\cos (\vec{q}_{2}\cdot \vec{r}+\phi)+\cos (\vec{q}_{3}\cdot \vec{r}-\phi))\tau_{y}].
\end{split}
\end{equation}
The first term $H_{\perp}$ in Eq. (\ref{eq:hperptr}) leads to the renormalization of the Dirac velocity \cite{LopesDosSantos2007,Bistritzer2011}. Therefore, $H_{0}$ and $H_{\perp}$ can be combined to produce $\tilde{H}_{0}$, which has the same form as $H_{0}$ but with the renormalized velocity $\tilde{v}_{F}$. The second term $H'$ in Eq. (\ref{eq:hperptr}) captures the $\theta$ dependence in $h(\pm \theta/2)$, which has been neglected in previous studies.

\emph{Dirac zero modes in a moir\'e supercell.---}The key observation in this work is that $H'$ is a spatially modulated Dirac mass term. It is periodic in the moir\'e superlattice and vanishes at two points in each supercell, which correspond to the AB and BA stacking regions. These regions form an emergent honeycomb lattice (see Fig. \ref{fig:lattice}). Moreover, there is a vortex or an antivortex structure around each zero point. For example, around $\vec{r}_{0}=(0,0)$ with an AB stacking pattern,
\begin{equation}
\tau(\vec{r})\simeq \frac{3}{4}w\theta k_{\theta}(x\tau_{y}-y\tau_{x}).
\end{equation}
Therefore, as a mapping from the real space to the space of Dirac mass terms, $\tau(\vec{r})$ has a vortex structure with vorticity $+1$ around $\vec{r}_{0}$. Similarly, around $\vec{r}_{1}=(4\sqrt{3}\pi/9k_{\theta},0)$ with a BA stacking pattern, $\tau(\vec{r})\simeq -\frac{3}{4}w\theta k_{\theta}((x-4\sqrt{3}\pi/9k_{\theta})\tau_{y}+y\tau_{x})$, thus there is an antivortex with vorticity $-1$.

Because these vortices are well separated in space, they can be treated independently at the first step. The Dirac Hamiltonian has a bound state solution with zero energy around each vortex. This is a general consequence of the celebrated index theorem of Dirac operators \cite{Jackiw1976, Jackiw1981, Bertlmann2000anomalies}. Here I give an explicit construction of the Dirac zero modes in the vortex background.

Let us first consider the vortex at $\vec{r}_{0}=(0,0)$. Denote $\tau(\vec{r})=\frac{3}{4}w\theta k_{\theta}f(r)(\tau_{y}\cos\theta_{r}-\tau_{x}\sin\theta_{r})$, in which $(r,\theta_{r})$ are the polar coordinates. The radial function $f(r) \sim r$ as $r\rightarrow 0$, and is assumed to approach a constant (denoted by $f_{\infty}$) of order $O(1/k_{\theta})$ as $r\rightarrow \infty$ \footnote{The precise asymptotic form of $f(r)$ is not significant because, as shown in the main text, the zero mode wavefunction at long distance is exponentially suppressed as $\tilde{v}_{F}$ vanishes at the magic angles.}. Let us find a zero-energy solution to the Dirac Hamiltonian $\tilde{H}_{0}+H'$.

Define the chirality operator $\chi=\sigma_{z}\otimes \tau_{z}$. It anticommutes with the Hamiltonian, hence all nonzero-energy eigenstates are paired up, i.e., an eigenstate $\Psi(\vec{r})$ with a nonzero energy $\epsilon\neq 0$ implies that $\chi\Psi(\vec{r})$ is an eigenstate with energy $-\epsilon$, and vice versa. However, the Atiyah-Singer index theorem guarantees that there is a unique unpaired normalizable zero-energy state with chirality $+1$ ($-1$) around a (anti)vortex with vorticity $+1$ ($-1$). Therefore, let us seek a zero-energy solution $\Psi_{0}(\vec{r})$ of the Dirac Hamiltonian in the form of $\Psi_{0}(\vec{r})=(\psi_{1}(\vec{r}),0,0,\psi_{4}(\vec{r}))^{T}$,
\begin{equation}
\tilde{v}_{F}i\vec{\partial}\cdot \vec{\sigma}^{*}\Psi_{0}(\vec{r}) +\sigma_{z}\otimes \tau(\vec{r})\Psi_{0}(\vec{r})=0.
\end{equation}
The explicit solution is given by $\psi_{1}(\vec{r})=\psi_{4}(\vec{r})=\psi(r)$, in which
\begin{equation}
\psi(r) = Ce^{-\frac{3w\theta k_{\theta}}{4\tilde{v}_{F}}\int_{0}^{r}f(r')dr'},
\end{equation}
and $C$ is the normalization constant.

Similarly, there is a unique unpaired zero mode with chirality $-1$ around the antivortex at $\vec{r}_{1}=(4\sqrt{3}\pi/9k_{\theta},0)$, which is given by $\Psi_{1}(\vec{r})=(0,\psi(|\vec{r}-\vec{r}_{1}|),\psi(|\vec{r}-\vec{r}_{1}|),0)^{T}$. There are eight zero modes in each moir\'e supercell if the spin and the valley degeneracies are taken into account.

The zero mode wavefunctions are exponentially localized around the vortex centers, $\psi(r)\sim Ce^{-r/\xi}$ as $r\rightarrow \infty$, in which $\xi =\frac{4\tilde{v}_{F}}{3w\theta k_{\theta}f_{\infty}}$ is the localization length. Remarkably, as $\tilde{v}_{F}$ vanishes at the magic angle, $\xi$ also vanishes and these zero modes are fully localized and separated from each other. Moreover, other eigenstates with nonzero energy are energetically separated from these zero modes due to the mass term by a characteristic energy scale $w\theta\simeq 2$ meV, which might be further enhanced due to the Coulomb interaction. Therefore, these zero modes are the most relevant degrees of freedom at low energy {\rr closest to the Fermi energy} at the magic angle, and should be responsible for the unconventional insulating and SC phases in TBG.

\emph{Moir\'e band formed by Dirac zero modes.---}The zero mode wavefunctions slightly overlap with each other in the vortex lattice. This induces effective hopping between zero modes on the emergent honeycomb lattice and leads to a moir\'e band structure. The effective Hamiltonian of this moir\'e band can be derived by projecting the Dirac Hamiltonian $\tilde{H}_{0}+H'$ into the subspace of zero modes. The result is a tight-binding model on the emergent honeycomb lattice,
\begin{equation} \label{eq:hteff}
H_{t,\mathrm{eff}}=-\sum_{ij}t_{ij}\Psi_{i}^{\dag}\Psi_{j},
\end{equation}
in which the hopping parameter $t_{ij}=\langle \Psi_{i}|\tilde{H}_{0}|\Psi_{j}\rangle$. $t_{ij}$ is exactly zero between any two zero modes with the same chirality because $\tilde{H}_{0}$ anticommutes with $\chi$, thus the effective hopping only takes place between zero modes with opposite chirality residing in different sublattices,
\begin{equation} \label{eq:tij}
t_{ij}\simeq 2ie^{i\theta_{ij}}\sqrt{4\pi}\tilde{v}_{F}\psi(|\vec{r}_{j}-\vec{r}_{i}|)
\end{equation}
if $\chi=+1$ at site $i$ and $-1$ at site $j$, in which $\theta_{ij}$ is the polar angle of $\vec{r}_{j}-\vec{r}_{i}$. {\rr This constraint of bipartite hopping originates from the particle-hole symmetry of the linearized massless Dirac fermions in the continuum approximation and should be relaxed in real material.}  $t_{ij}$ decays exponentially with $|\vec{r}_{j}-\vec{r}_{i}|$, thus only the nearest-neighbor hopping will be retained. The total phase of the hopping $t_{ij}$ accumulated around each hexagonal plaquatte is zero, thus the phase factors in $t_{ij}$ can be eliminated by a gauge transformation. Therefore, this moir\'e band is described by a zero-flux tight-binding model on the honeycomb lattice. {\rr The upper edge of this moir\'e band may merge into the high-energy states if the hopping amplitude is comparable to the Dirac mass $w\theta$.} Furthermore, the amplitude of $t_{ij}$ is exponentially suppressed as $\tilde{v}_{F}$ vanishes at the magic angle due to the $\tilde{v}_{F}$ dependence in the $\psi(r)$ factor, thus this low-energy moir\'e band becomes extremely flat and strongly correlated at the magic angle.

\emph{Mott insulating phase and beyond.---}Retrieving the spin and the valley degrees of freedom and including the electron Coulomb interaction, the full effective Hamiltonian of the low-energy moir\'e band is the following Hubbard model,
\begin{equation} \label{eq:hubbard}
H_{\mathrm{eff}}=-t\sum_{\alpha=1}^{4}\sum_{\langle ij\rangle}\Psi_{i\alpha}^{\dag}\Psi_{j\alpha}+\mathrm{h.c.} +\frac{1}{2}U\sum_{i}n_{i}(n_{i}-1),
\end{equation}
in which $\alpha$ denotes the combined spin and valley index. $n_{i}=\sum_{\alpha}\Psi_{i\alpha}^{\dag}\Psi_{i\alpha}$ is the total electron number at site $i$. The onsite Coulomb repulsion is estimated by $U\simeq e^{2}/4\pi\varepsilon_{0}\varepsilon\lambda\simeq 30~\mathrm{meV}$, in which $\varepsilon\simeq 3$ is the relative dielectric constant of the hexagonal BN substrate \cite{Kim2017, Cao2018}.

This Hubbard model has an emergent SU(4) symmetry in the combined spin-valley space. At $1/4$ or $3/4$ filling, i.e., $\pm 2$ excess electrons per supercell away from the charge neutral point, the strong Coulomb repulsion, $U\gg t$, leads to a Mott insulating phase. The Hubbard interaction favors spin-valley singlet states. This is captured by the following spin-valley exchange interaction, which is also called the Kugel-Khomskii model \cite{Kugel1982},
\begin{equation} \label{eq:hex}
H_{J}=J\sum_{\langle ij\rangle}(\vec{S}_{i}\cdot \vec{S}_{j}+1/4)(\vec{T}_{i}\cdot \vec{T}_{j}+1/4),
\end{equation}
with $J=8t^{2}/U$. $\vec{S}_{i}$ and $\vec{T}_{i}$ are the SU(2) generators acting on the spin and the valley spaces, respectively.

The spin-valley exchange interaction Eq. (\ref{eq:hex}) has been shown to lead to a quantum spin-orbital (valley) liquid at $1/4$ filling, i.e., a quantum disordered phase without any spontaneous symmetry breaking \cite{Corboz2012}. Slightly doped away from $1/4$ or $3/4$ filling, the system is captured by the SU(4)-symmetric $t$-$J$ model, and may become superconducting at the ground state. The quantum spin-orbital liquid phase and its adjacent SC phase will be explored in detail in future works.

\emph{Summary and discussions.---}In summary, I have shown that the interlayer hopping in the twisted bilayer graphene gives rise to a spatially modulated Dirac mass term in the continuum model. There are two Dirac zero modes localized in each moir\'e supercell due to the vortex lattice structure of the mass term. These zero modes form a moir\'e band in this emergent honeycomb vortex lattice. This band becomes extremely flat at the magic angle in particular, thus the Coulomb interaction predominates and leads to a quantum disordered Mott insulating phase at $1/4$ or $3/4$ filling, which may account for the unconventional insulating phases discovered in experiments.

This model is justified by the fact that the localization length of the zero modes and the moir\'e band width are exponentially suppressed as the Dirac velocity vanishes at the magic angle. Therefore, this moir\'e band constitutes the lowest-energy electronic states and should be responsible for the unconventional phases in magic-angle twisted bilayer graphene.

Both the spin and the valley degeneracies are maintained in this effective model because the valley-mixing term in the interlayer hopping has been neglected due to the small momentum transfer \cite{Bistritzer2011}. However, if this effect is taken into account, which might be further enhanced by the strong correlation, the valley degeneracy can be lifted at low energy. This may account for the two-fold (instead of four-fold) degeneracy extracted from the quantum oscillations \cite{Cao2018a}. A detailed analysis of the valley-mixing effect and especially its impact on the unconventional superconductivity is left for future works.

\acknowledgments
I am indebted to Wenxin Ding for bringing my attention to this issue and many enlightening discussions. This work is supported by National Key R\&D Program of China (No. 2018YFA0305800), National Natural Science Foundation of China (No. 11804337), Strategic Priority Research Program of CAS (No. XDB28000000), and Beijing Municipal Science \& Technology Commission (No. Z181100004218001).

\bibliographystyle{apsrev4-1_custom}
\bibliography{library}
\end{document}